\documentstyle[12pt,epsf]{article}

 \def\lsi{\raise0.3ex\hbox{$<$\kern-0.75em\raise-1.1ex\hbox{$\sim$}}}
 \def\gsi{\raise0.3ex\hbox{$>$\kern-0.75em\raise-1.1ex\hbox{$\sim$}}}
 
 \newcommand{\gsim}{\mathop{\gsi}}

\begin{document}

\begin{titlepage}

\hfill hep-lat/0112001

\begin{centering}
\vfill

{\bf ANTISYMMETRIC AND OTHER SUBLEADING CORRECTIONS \\
TO SCALING IN THE LOCAL POTENTIAL APPROXIMATION
}

\vspace{1cm}

M.~M.~Tsypin\footnote{Also at Lebedev Physical Institute, Moscow

\hspace*{0.5em} E-mail: tsypin@ias.edu}

\vspace{1cm}

{\em Department of Physics and Astronomy, Rutgers University, \\ 
Piscataway, NJ 08854, USA}

\vspace{0.3cm}

and

\vspace{0.3cm}

{\em Institute for Advanced Study, Einstein Drive,
Princeton, NJ 08540, USA}

\vspace{2cm}
{\bf Abstract}

\end{centering}

\vspace{0.3cm}\noindent

For systems in the universality class of the three-dimensional
Ising model we compute the critical exponents in the local potential
approximation (LPA), that is, in the framework of the
Wegner-Houghton equation. We are mostly interested in antisymmetric
corrections to scaling, which are relatively poorly studied.
We find the exponent for the leading antisymmetric correction
to scaling $\omega_A \approx 1.691$ in the LPA. This high value
implies that such corrections cannot explain asymmetries 
observed in some Monte Carlo simulations.

\vfill \vfill

\end{titlepage}

\section{Introduction}
This work is devoted to the study of corrections to scaling in the
vicinity of the critical point: the endpoint of the line of first 
order phase transition.

There are many systems that have a critical point belonging to the 
universality class of the three-dimensional (3D) Ising model.
In easy-axis magnetic materials that exhibit a phase transition
between paramagnetic and ferromagnetic phases (these systems
most directly correspond to the Ising model) the first order
phase transition line is in the plane $(T,h)$, where $T$ is the
temperature and $h$ is external magnetic field. It
is located at $h=0$ and $T<T_c$, and ends at the critical
point $(T=T_c,h=0)$. 

The system liquid-gas has the line of 
first order transition, ending at the critical point, in the 
plane (temperature, pressure). This critical point has been
very thoroughly studied. A comprehensive review, as well
as additional references, can be found in~\cite{PeVi00}.
Experimental data, theoretical considerations~\cite{NZ81,RehrMermin}
and Monte Carlo simulations of model systems~\cite{BruWil}
all indicate that it belongs to the 3D Ising model universality 
class.

The Standard Model of electroweak interactions was shown to 
have a similar phase transition line in the plane (higgs mass,
temperature)~\cite{KirzhLin,basicEWPT}. 
Monte Carlo simulations provide convincing
evidence that the endpoint of this line belongs to the 3D Ising
universality class~\cite{ourEWPT}.

There are strong arguments in favor of a similar first order phase
transition line in hot nuclear matter, described by quantum
chromodynamics (QCD) at high temperature, in this case in the
plane (chemical potential for baryons, 
temperature)~\cite{Misha,FodorKatz}. Its endpoint is conjectured 
to belong to the 3D Ising universality class~\cite{Misha}.

Monte Carlo simulations play an important role
in elucidating the critical properties of these systems, especially 
of more complex ones, such as electroweak matter and nuclear matter 
at high temperature. These simulations normally have to be performed 
for systems of relatively small sizes, usually no larger than 
$64^3$~\cite{ourEWPT} (and often considerably smaller), due to
their high complexity and the corresponding high computational load.
This means that the system is not very deep into the scaling region
(roughly speaking, one cannot reach the correlation length $\xi$
much larger than the system size $L$), and the interpretation of
the Monte Carlo data requires proper account for corrections
to scaling.

Here we come to the following interesting point. Unlike the
Ising model, phase transitions in the liquid-gas system, 
as well as in electroweak and nuclear matter, do not have
the exact global $Z_2$ symmetry that corresponds to changing
the sign of magnetization in the Ising model, i.~e.\ to 
simultaneous flipping of all spins. For example, in the
Ising model (without external field) the probability distribution
for the total magnetization of the system is always perfectly
symmetric, despite corrections to scaling, while this is not
necessarily so in above-mentioned models. The $Z_2$ symmetry
is expected to be dynamically restored in the scaling limit~\cite{NZ81},
but there can be, in addition to the usual corrections to scaling,
which are even in the order parameter, also corrections that are odd.

Monte Carlo studies of critical points in such systems,
which typically aim at obtaining the linear mapping of the vicinity
of the critical point in the plane of parameters of the model onto
the corresponding vicinity of the critical point of the 3D Ising
model in the $(T,h)$ plane, have to somehow resolve the issue
of deviations from the $Z_2$ symmetry~\cite{BruWil,ourEWPT}. 
Suffice it to say
that such mapping requires determination of the magnetization-like
($M$-like) and energy-like ($E$-like) directions in the space
of observables, which would be much easier in the presence of
$Z_2$ symmetry: the requirement that the probability distribution
of the $M$-like observable should be symmetric greatly aids the
search for the corresponding direction in the space of 
observables~\cite{BruWil}.

In practice, the data produced in Monte Carlo simulations demonstrate
quite non-negligible deviations from $Z_2$ symmetry, and an interesting 
question is whether these deviations can be attributed to antisymmetric,
i.~e.\ $M$-odd, corrections to scaling~\cite{ourEWPT}. Such corrections
have attracted much less attention in the literature than the usual
($M$-even) corrections; however, several studies in the framework
of the $\epsilon$-expansion~\cite{Wegner,NZ81,ZhangZia}, as well
as renormalization group~\cite{NewRie}, have been published.

While the usual $M$-even corrections to scaling behave as
$L^{-\omega},\ \omega \equiv \Delta/\nu \approx 0.8$, $L$ being
the characteristic length in the system, the $M$-odd corrections
to scaling are governed by their own exponent $\omega_A$
(we prefer this notation~\cite{PeVi00} to traditional $\omega_5$).
The $\epsilon$-expansion obtained in~\cite{ZhangZia} reads
\begin{equation}
\omega_A = 1 + {11 \over 6} \epsilon -
            {685 \over 324} \epsilon^2 +
            {107855 + 103680 \zeta(3) \over 34992} \epsilon^3
            + O(\epsilon^4).
\end{equation}
At $\epsilon = 1$ this series behaves too poorly 
$(1+1.83-2.11+6.64\ \ldots)$ to get conclusive results;
Pad\'e approximants produce the sequence 2.83, 1.85, 2.32 in 
orders $\epsilon$, $\epsilon^2$, $\epsilon^3$, respectively,
leading to the estimate $\omega_A \gsim 1.5$~\cite{ZhangZia}.
The renormalization group computation~\cite{NewRie} gives
$\omega_A = 2.4(5)$.

These very high values of $\omega_A$ imply that $M$-odd
corrections to scaling, going as $L^{-\omega_A}$, should decay
much faster than the leading $M$-even corrections $(L^{-\omega})$,
and quickly become negligible when the scaling limit is approached.
However, in the Monte Carlo study~\cite{ourEWPT} the asymmetry
of the probability distribution of the $M$-like observable
at the critical point was found to decay much more slowly 
with the growing lattice size, more like $L^{-0.5}$.

Before concluding that this observation requires explanation
outside the scope of corrections to scaling, one has to be confident
that there are indeed no $M$-odd corrections to scaling with
small exponents (which could conceivably happen, for example,
if existing computations of $\omega_A$ somehow missed the leading
$M$-odd correction and corresponded instead to a subleading correction).
To gain such confidence, we have performed a computation of 
$M$-odd corrections in the framework of the local potential
approximation (LPA), i.~e.\ the Wegner-Houghton 
equation~\cite{WH,HH,BB01,BTW}.

The WH equation satisfies our needs nicely, as it reproduces robustly
all the relevant features of the theory. In this approximation
the critical exponent $\eta$ is zero. 

Our study differs from the existing literature on the WH 
equation~\cite{WH,HH,PR,BB90,CT,BB2000} in several significant
aspects: (1) we are not aware of any previous study of $M$-odd
corrections to scaling in the framework of the WH equation,
with the exception of unpublished work~\cite{BBS}; 
(2) we do not rely on avoiding the singularity when solving the 
WH equation, and thus check whether the large-field domain
plays an important role in fixing the values of the critical
exponents.

Our main results are as follows. (1) We confirm the absence of 
$M$-odd corrections to scaling with small exponents. We get
the exponent for the leading $M$-odd correction $\omega_A \approx 1.691$,
which is consistent with~\cite{BBS}. (2) The large-field domain
is not important for fixing the critical exponents, at our 
level of accuracy.

We conclude that the asymmetry observed in~\cite{ourEWPT} is not
explainable by $M$-odd corrections to scaling.

The paper is organized as follows. In Sect.\ 2 we discuss the local
potential approximation and the Wegner-Houghton equation. 
In Sect.\ 3 we describe our method of finding the fixed point and 
computing the critical exponents. Sect.\ 4 contains our numerical
results and conclusions.

\section{The Wegner-Houghton equation}
In this section we remind the reader the structure and the meaning
of the WH equation. Accurate derivation can be found in original
papers~\cite{WH,HH} and reviews~\cite{BB01,BTW}.

The starting point is the description of the system in the vicinity
of the critical point by the theory of the 1-component real scalar
field $\phi$ with the bare potential $V(\phi)$:
\begin{eqnarray}
  \cal{Z} & = & \int D \phi e^{-S[\phi]}, \\
  S & = & \int d^d x
    \left\{ 
    {1 \over 2} Z \partial_\mu \phi \partial_\mu \phi + V(\phi)
    \right\} .
\end{eqnarray}
We will be eventually interested in the three-dimensional case $(d=3)$.
The effective potential $U_0(\varphi)$ in one-loop approximation reads
\begin{equation}
 U_0(\varphi) = V(\varphi) + 
  {1 \over 2} \int_{|q| < \Lambda} \frac{d^d q}{(2\pi)^d} 
  \ln \left( Zq^2 + \frac{d^2 V}{d \varphi^2} \right) . 
\end{equation}
Let us modify the region of integration in the right hand side 
from all $q$ below the cutoff $\Lambda$ to $q$ above certain
momentum $k$ and below $\Lambda$. This will produce the $k$-dependent
analog of the effective potential, denoted by $U_k$:
\begin{eqnarray}
 U_k(\varphi) & = & V(\varphi) + 
  {1 \over 2} \int_{k< |q| < \Lambda} \frac{d^d q}{(2\pi)^d} 
  \ln \left( Zq^2 + \frac{d^2 V}{d \varphi^2} \right)   \\
  & = & V(\varphi) + 
  {K_d \over 2} \int_k^\Lambda q^{d-1} dq 
  \ln \left( Zq^2 + \frac{d^2 V}{d \varphi^2} \right) , 
\end{eqnarray}
where we have denoted by $K_d$ the area of a $d$-dimensional unit
sphere, divided by $(2\pi)^d$. For $d=3$, $K_3 = 1/(2\pi^2)$.
Taking derivative over $k$, subtracting the $\varphi$-independent
quantity $ - {1 \over 2} K_d k^{d-1} \ln k^2 $, and replacing $V(\varphi)$
in the right hand side with $U_k(\varphi)$ (``renormalization group
improvement''), we obtain
\begin{equation}
 \partial_k U_k(\varphi)  =  
  {K_d \over 2} k^{d-1}  
  \ln \left( Z + 
  {1 \over k^2} \frac{\partial^2 U_k}{\partial \varphi^2} \right) .
  \label{dkUk}
\end{equation}
This equation describes the evolution of the scale-dependent effective
potential $U_k(\varphi)$ with the change of scale $k$. For the purposes
of the study of the vicinity of the critical point, it is convenient
to convert it to the form that includes rescaling of $\varphi$ and
$U_k$. We introduce the dimensionless parameter $t$: $k = \Lambda e^{-t}$,
and rescaled quantities $\tilde{U}_t$ and $\tilde{\varphi}_t$:
$U_k = k^d \tilde{U}_t$, $\varphi = k^{d_\varphi} \tilde{\varphi}_t$,
where $d=3$ is the dimensionality of space, and $d_\varphi$ is
the dimension of the field $\varphi$. In terms of $t$, $\tilde{U}_t$ and 
$\tilde{\varphi}_t$ eq.~(\ref{dkUk}) reads
\begin{equation}
 {\partial \tilde{U}_t \over \partial t}   =  
  {K_d \over 2}
  \ln \left( Z + k^{d-2-2d_\varphi} 
  \frac{\partial^2 \tilde{U}_t }{\partial \tilde{\varphi}_t^2} \right) +
  d \cdot \tilde{U}_t - 
  d_\varphi \cdot \tilde{\varphi}_t 
  \frac{\partial \tilde{U}_t }{\partial \tilde{\varphi}_t} .
\end{equation}
Using the canonical dimension of $\varphi$: $d_\varphi = (d-2)/2$,
dropping all the tildes and $t$-subscripts, and denoting derivatives
over $\varphi$ by primes,
\begin{equation}
  {\partial U \over \partial t} = 
  {K_d \over 2} \ln ( Z + U'' ) + 
  d \cdot U - {d-2 \over 2} \varphi U' .
\end{equation}
As in this approximation $Z$ is a constant that depends neither on 
$\varphi$ nor on $t$, one can conveniently set $Z=1$. Restricting to
$d=3$, we finally get
\begin{equation}
  {\partial U \over \partial t} = 
  {1 \over 4 \pi^2} \ln (1 + U'') + 
  3 U - {1 \over 2} \varphi U' .      \label{flow3d}
\end{equation}
This is the form of the WH equation that we will use in the following.
(One should keep in mind that is does not treat properly the 
additive constant, that is, the $\varphi$-independent part of $U$).

\section{Fixed point and critical exponents}
Equation~(\ref{flow3d}) describes the evolution of the effective
potential $U(\varphi)$ with the change of scale. Let us denote
its right hand side by $F(\varphi)$. The fixed point is described 
by effective potential that does not evolve with $t$, that is,
$U_\ast(\varphi)$ such that $F(\varphi) = const$. Usually one finds
$U_\ast(\varphi)$ by numerically solving the differential equation
$F'(\varphi) = 0$ and fixing the solution by requiring that it
does not run into singularity at large $\varphi$~\cite{HH,CT,BB2000,BB01}.
Analysis of the evolution over $t$ of small deviations from 
$U_\ast(\varphi)$ produces the critical exponents.

\begin{figure}
   \centerline{ \epsfxsize=4in\epsfbox{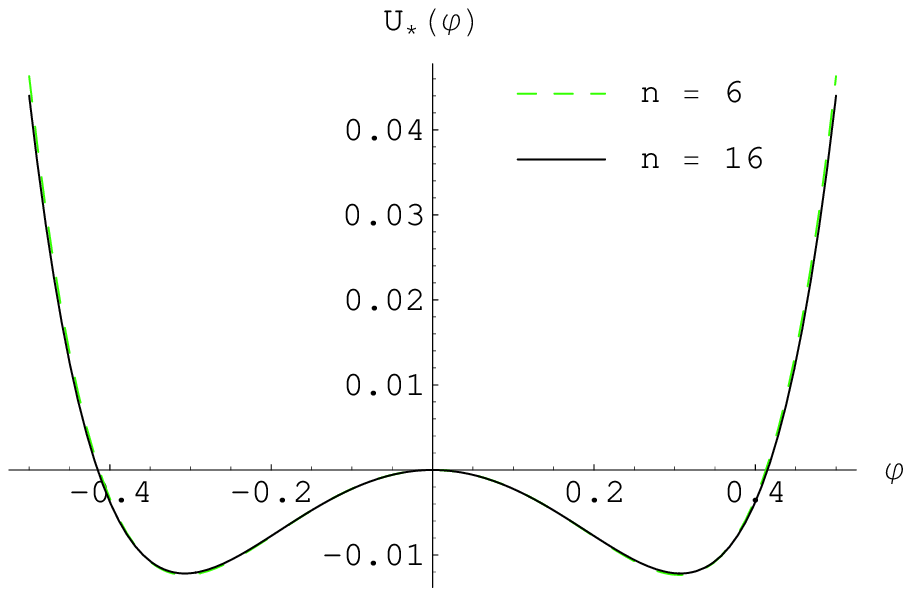} }
   \centerline{ \epsfxsize=4in\epsfbox{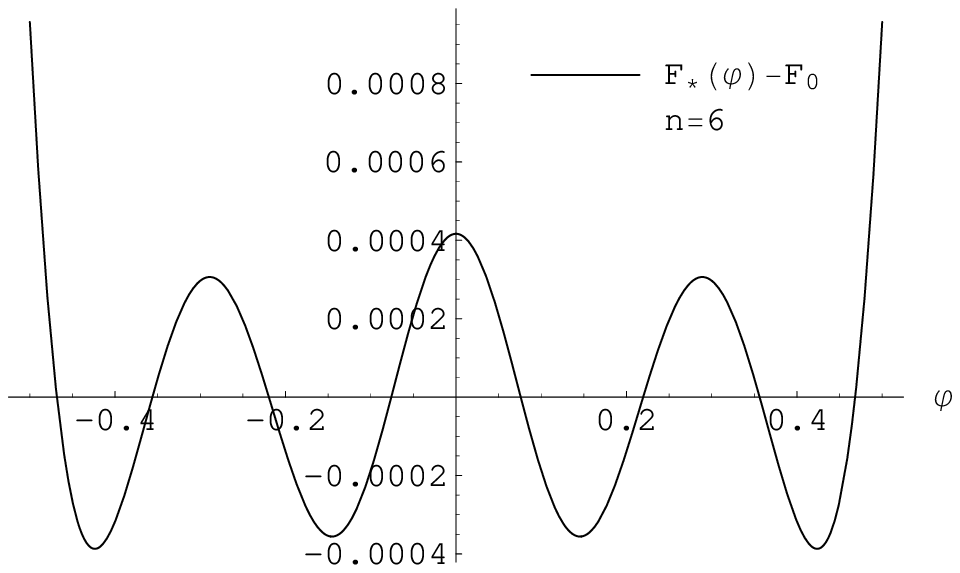}    }
   \centerline{ \epsfxsize=4in\epsfbox{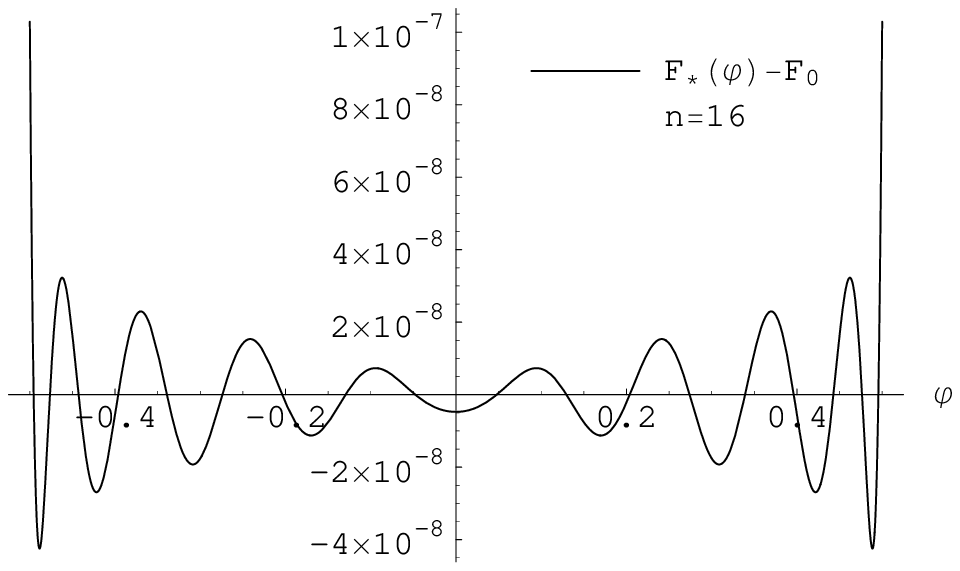}   }
\caption{
Approximate fixed point effective potential $U_\ast(\varphi)$ and the
corresponding deviation of the right hand side of 
eq.~(\protect\ref{flow3d}) from constant, for $n=6$ and $n=16$
($\varphi_{\rm max} = 0.5$ in both cases; the two graphs for 
$U_\ast(\varphi)$ are hardly distinguishable).
 }
\label{fig1}
\end{figure}

Our approach to computation of $U_\ast(\varphi)$ and critical exponents
is as follows. We always approximate $U(\varphi)$ (both $U_\ast(\varphi)$
and $U_\ast(\varphi)$ + perturbations) by polynomials up to a certain
order $n$ in $\varphi$:
\begin{equation}
  U_\ast(\varphi) = \sum_{j=1}^n a_j \varphi^j .    \label{Ustar}
\end{equation}
To determine $U_\ast(\varphi)$, we search for parameters 
$a_2, a_4, \ldots a_n$ of expansion~(\ref{Ustar}) that minimize
the deviation of $F(\varphi)$ from constant on a reasonably
chosen interval. That is, we minimize
\begin{equation}
  \sigma^2 = {1 \over 2 \varphi_{\rm max}}
  \int_{-\varphi_{\rm max}}^{\varphi_{\rm max}}
  \Bigl( F(\varphi) - F_0 \Bigr)^2 d \varphi .      \label{sigma}
\end{equation}
At this stage the odd coefficients of expansion~(\ref{Ustar}) are zero, 
and minimization involves $n/2$ even coefficients of~(\ref{Ustar}), plus
$F_0$. The achievable proximity of $F(\varphi)$ to constant is
improving rapidly with the increasing order of the approximation
(Fig.~\ref{fig1}). 

The next step is the study of deviations of $U(\varphi)$ from 
$U_\ast(\varphi)$, and their evolution over $t$. As will be 
clear below, it is convenient to introduce a separate set of
parameters, $b_i = b_i^\ast + \delta b_i$, $i=1 \ldots n$, 
to parameterize the deviation of $U$ from $U_\ast$ (which 
corresponds to $b_i^\ast$). Then the evolution of parameters
follows from
\begin{equation}
  {\partial \over \partial t} U(b_i^\ast + \delta b_i) = 
  F(b_i^\ast + \delta b_i) - F_\ast .
\end{equation}
We have to subtract $F_\ast \equiv F(b_i^\ast)$ in the right hand
side, to account for the approximate nature of our treatment 
of the fixed point. Thus
\begin{equation}
  {\partial U \over \partial b_j} {d \over dt} \delta b_j = 
  {\partial F \over \partial b_j} \delta b_j .
\end{equation}
Here $\partial U / \partial b_j$ and $\partial F / \partial b_j$
are understood to be taken at the fixed point, $b_i^*$.
Defining the scalar product
\begin{equation}
  \langle f_1(\varphi) | f_2(\varphi) \rangle = 
  {1 \over 2 \varphi_{\rm max}}
  \int_{-\varphi_{\rm max}}^{\varphi_{\rm max}}
   f_1(\varphi) f_2(\varphi) d \varphi ,
\end{equation}
and introducing matrices 
\begin{equation}
  A_{ij} = \biggl\langle {\partial U \over \partial b_i} \biggl|
   {\partial U \over \partial b_j} \biggr\rangle,
  \quad
  B_{ij} = \biggl\langle {\partial U \over \partial b_i} \biggl|
   {\partial F \over \partial b_j} \biggr\rangle,
\end{equation}
we obtain
\begin{equation}
  A \cdot {d \over dt} \vec{\delta b} = B \cdot \vec{\delta b} .
\end{equation}
Thus
\begin{equation}
  \vec{\delta b} = \sum_{i=1}^n c_i e^{\lambda_i t} \vec{v}_i ,
\end{equation}
where $\lambda_i$ and $\vec{v}_i$ are, respectively, eigenvalues and
eigenvectors of $A^{-1}B$. $\lambda_i$ are exactly the critical exponents
we are interested in.

To simplify the computation and improve its numerical stability, we
parameterize the deviation of $U$ from $U_\ast$ so that 
$A_{ij} = \delta_{ij}$, namely,
\begin{equation}
  U(\varphi) = U_\ast(\varphi) + \sum_{j=1}^n b_j \tilde{P}_j(\varphi) ,
\end{equation}
where $\tilde{P}_j(\varphi)$ are Legendre polynomials $P_j$ normalized
on $[ -\varphi_{\rm max}, \varphi_{\rm max} ]$:
\begin{equation}
  \tilde{P}_j(\varphi) = \sqrt{2j+1 \over 2\varphi_{\rm max}}
    P_j \left( { \varphi \over \varphi_{\rm max} } \right).
\end{equation}
Then $\lambda_i$ are just the eigenvalues of the matrix $B$.

\section{Results}
The results for critical exponents are collected in Tables 1--3.
We use polynomial approximations of order $6 \ldots 16$, for three
values of $\varphi_{\rm max}$: 0.5, 0.44 and 0.4. We observe that
the higher eigenvalues are almost insensitive to the choice of
$\varphi_{\rm max}$. This is fortunate, and provides additional
evidence that the LPA captures correctly the important part
of physics. The usual approach relies on avoiding the singularity
at large $\varphi$~\cite{HH,CT,BB2000,BB01}, and one may wonder 
to what extent the results are determined by the asymptotical
properties of the WH equation at large $\varphi$, rather than
by properties at physically relevant range of $\varphi$
(of order of the position of the minimum of $U_\ast$).

Our values of the critical exponents in the even sector,
\begin{equation}
  y_t \approx 1.45041, \ \omega \approx 0.5952, \ 
  \omega_2 \approx 2.838, \ \omega_3 \approx 5.18,
\end{equation}
are in agreement with the most accurate of the previous 
computations~\cite{CT}:
\begin{equation}
  \nu \approx 0.6895, \ \omega \approx 0.5952, 
  \lambda_2 \approx -2.8384, \lambda_3 \approx -5.1842.
\end{equation}
The most interesting part of our results is the set of critical exponents
for the odd sector:
\begin{equation}
  y_h = 2.5, \ y_{\rm shift} = 0.5, \ 
  \omega_A \approx 1.691, \ \omega_{A2} \approx 4.00.
\end{equation}
The appearance of the somewhat unusual exponent $y_{\rm shift}$ 
reflects the fact that, strictly speaking, in the case of asymmetric
$U(\varphi)$ the renormalization group transformation should include,
in addition to blocking and rescaling of $\varphi$, also the shift 
of $\varphi$. As we did not take this into account, this additional
exponent emerged.

The value of $\omega_A$ is consistent with~\cite{BBS} but has higher 
precision. We are not aware of any previous computation of $\omega_{A2}$.

To summarize, we have computed the critical exponents, in $\varphi$-even
as well as in $\varphi$-odd sectors, for the systems in the 3D Ising
universality class, in the local potential approximation. We show
that the previously known values of the critical exponents in the 
$\varphi$-even sector are reproduced, within our accuracy, even without 
relying upon avoiding the singularity in the Wegner-Houghton equation 
at $\varphi \to +\infty$. The absence of slowly-decaying $\varphi$-odd
corrections to scaling implies that they are not responsible
for asymmetries observed in Monte Carlo studies~\cite{ourEWPT}.

I am deeply grateful to Prof.\ H.~Neuberger for his kind hospitality,
support, and many interesting discussions. I would like to thank
the Department of Physics and Astronomy, Rutgers University
and the School of Natural Sciences, IAS, for their hospitality 
and support. This work was supported in part by DOE grants
DE-FG02-96ER40949 and DE-FG02-90ER40542.

\newpage

\begin{table}
{%\small
 \begin{center}
  \begin{tabular}{|llllllll|}
   \hline
$n$  & 
    & 16        & 14        & 12        & 10        & 8         & 6         \\
$\sigma$ &
    & ${\scriptstyle 1.48 \times 10^{-8} }$
    & ${\scriptstyle 2.36 \times 10^{-7} }$
    & ${\scriptstyle 8.48 \times 10^{-7} }$
    & ${\scriptstyle 2.31 \times 10^{-6} }$
    & ${\scriptstyle 4.13 \times 10^{-5} }$
    & ${\scriptstyle 2.78 \times 10^{-4} }$
    \\ 
   \hline
$y_h$  & $A$
    & 2.5       & 2.5       & 2.5       & 2.5       & 2.5       & 2.5       \\
$y_t$  & $S$
    & 1.450413  & 1.450412  & 1.450402  & 1.45044   & 1.44878   & 1.44099   \\
$y_{\rm shift}$  & $A$
    & 0.5       & 0.500003  & 0.500035  & 0.500223  & 0.505546  & 0.491122  \\
$-\omega$ & $S$
    & -0.595235 & -0.595256 & -0.595119 & -0.595044 & -0.589349 & -0.530546 \\
$-\omega_A$ & $A$
    & -1.69132  & -1.6914   & -1.69099  & -1.70327  & -1.5294   & -0.920653 \\
$-\omega_2$ & $S$
    & -2.83838  & -2.83869  & -2.83794  & -2.89125  & -2.50668  & -2.00335  \\
$-\omega_{A2}$ & $A$
    & -3.99828  & -4.00259  & -4.11481  & -3.46038  & -3.38734  &           \\
$-\omega_3$  & $S$
    & -5.18418  & -5.20558  & -5.37167  & -4.48014  & -5.0081   &           \\
 & $A$
    & -6.44866  & -6.64226  & -5.61029  & -6.89618  &           &           \\
 & $S$  
    & -7.76251  & -7.95578  & -6.91652  & -9.08612  &           &           \\
 & $A$  
    & -9.35137  & -8.46459  & -11.5582  &           &           &           \\
 & $S$  
    & -10.8738  & -10.2681  & -14.3183  &           &           &           \\
 & $A$  
    & -12.3235  & -17.3682  &           &           &           &           \\
 & $S$  
    & -14.6305  & -20.7111  &           &           &           &           \\
 & $A$  
    & -24.3498  &           &           &           &           &           \\
 & $S$  
    & -28.2887  &           &           &           &           &           \\
   \hline
  \end{tabular}
 \end{center}
}
\caption{
Critical exponents for $\varphi_{\rm max} = 0.5$. 
The quality of approximation is characterized by $\sigma$, 
according to~(\protect\ref{sigma}). Second column indicates
the symmetry of the eigenvector ($S$ = even, $A$ = odd).
} \label{table1}

\end{table}

\begin{table}
{%\small
 \begin{center}
  \begin{tabular}{|llllllll|}
   \hline
$n$  & 
    & 16        & 14        & 12        & 10        & 8         & 6         \\
$\sigma$ &
    & ${\scriptstyle 3.41 \times 10^{-9} }$
    & ${\scriptstyle 3.96 \times 10^{-8} }$
    & ${\scriptstyle 3.14 \times 10^{-7} }$
    & ${\scriptstyle 3.43 \times 10^{-7} }$
    & ${\scriptstyle 1.41 \times 10^{-5} }$
    & ${\scriptstyle 1.28 \times 10^{-4} }$
    \\ 
   \hline
$y_h$  & $A$
    & 2.5       & 2.5       & 2.5       & 2.5       & 2.5       & 2.5       \\
$y_t$  & $S$
    & 1.450410  & 1.45035   & 1.45056   & 1.4505    & 1.44882   & 1.45284   \\
$y_{\rm shift}$  & $A$
    & 0.499991  & 0.499918  & 0.500252  & 0.500054  & 0.498699  & 0.523208  \\
$-\omega$ & $S$
    & -0.595165 & -0.595677 & -0.594694 & -0.593543 & -0.604152 & -0.546673 \\
$-\omega_A$ & $A$
    & -1.69158  & -1.69361  & -1.68492  & -1.7014   & -1.75849  & -1.92471  \\
$-\omega_2$ & $S$
    & -2.83969  & -2.84017  & -2.83152  & -2.90894  & -3.08336  & -3.59606  \\
$-\omega_{A2}$ & $A$
    & -3.99407  & -4.03145  & -4.2279   & -4.56781  & -5.65287  &           \\
$-\omega_3$  & $S$
    & -5.19053  & -5.32954  & -5.73614  & -6.33275  & -8.01733  &           \\
 & $A$
    & -6.77786  & -7.47611  & -8.36842  & -10.7213  &           &           \\
 & $S$  
    & -8.44641  & -9.46912  & -10.6988  & -13.8132  &           &           \\
 & $A$  
    & -11.7329  & -13.3336  & -17.2785  &           &           &           \\
 & $S$  
    & -14.2826  & -16.2727  & -21.1253  &           &           &           \\
 & $A$  
    & -19.5243  & -25.361   &           &           &           &           \\
 & $S$  
    & -23.0967  & -29.9828  &           &           &           &           \\
 & $A$  
    & -34.9966  &           &           &           &           &           \\
 & $S$  
    & -40.4086  &           &           &           &           &           \\
   \hline
  \end{tabular}
 \end{center}
}
\caption{
Critical exponents for $\varphi_{\rm max} = 0.44$.
} \label{table2}

\end{table}

\begin{table}
{%\small
 \begin{center}
  \begin{tabular}{|llllllll|}
   \hline
$n$  & 
    & 16        & 14        & 12        & 10        & 8         & 6         \\
$\sigma$ &
    & ${\scriptstyle 1.65 \times 10^{-9} }$
    & ${\scriptstyle 6.53 \times 10^{-9} }$
    & ${\scriptstyle 1.15 \times 10^{-7} }$
    & ${\scriptstyle 3.14 \times 10^{-7} }$
    & ${\scriptstyle 5.78 \times 10^{-6} }$
    & ${\scriptstyle 6.64 \times 10^{-5} }$
    \\ 
   \hline
$y_h$  & $A$
    & 2.5       & 2.5       & 2.5       & 2.5       & 2.5       & 2.5       \\
$y_t$  & $S$
    & 1.45031   & 1.45014   & 1.45136   & 1.45019   & 1.44437   & 1.47603   \\
$y_{\rm shift}$  & $A$
    & 0.499857  & 0.499766  & 0.501045  & 0.499452  & 0.493915  & 0.537914  \\
$-\omega$ & $S$
    & -0.595091 & -0.597076 & -0.593018 & -0.590154 & -0.619136 & -0.6057   \\
$-\omega_A$ & $A$
    & -1.69522  & -1.6978   & -1.66645  & -1.73619  & -2.07471  & -2.76637  \\
$-\omega_2$ & $S$
    & -2.84795  & -2.83693  & -2.82088  & -3.05967  & -3.72532  & -4.91745  \\
$-\omega_{A2}$ & $A$
    & -3.98292  & -4.09967  & -4.59586  & -5.64517  & -7.68927  &           \\
$-\omega_3$  & $S$
    & -5.20884  & -5.56624  & -6.46     & -7.98584  & -10.7612  &           \\
 & $A$
    & -7.30803  & -8.68827  & -10.7014  & -14.2306  &           &           \\
 & $S$  
    & -9.3943   & -11.2751  & -13.8281  & -18.2123  &           &           \\
 & $A$  
    & -14.2416  & -17.3769  & -22.65    &           &           &           \\
 & $S$  
    & -17.6035  & -21.3259  & -27.5739  &           &           &           \\
 & $A$  
    & -25.6903  & -32.9993  &           &           &           &           \\
 & $S$  
    & -30.4805  & -38.8901  &           &           &           &           \\
 & $A$  
    & -45.2637  &           &           &           &           &           \\
 & $S$  
    & -52.1331  &           &           &           &           &           \\
   \hline
  \end{tabular}
 \end{center}
}
\caption{
Critical exponents for $\varphi_{\rm max} = 0.4$. 
} \label{table3}

\end{table}

\end{document}